**Compensation rule between the parameters of the Jonscher's Universal Dielectric Response in disordered materials**

**Anthony N. Papathanassiou, Elias Sakellis**

***National and Kapodistrian University of Athens, Department of Physics,***
***Section of Condensed Matter Physics,***
***Panepistimioplois, 15784 Zografos, Athens, Greece***

## Abstract

Experimental results for a huge number of different types of disordered materials published during the past fifty years confirm the validity of the Jonscher's Universal Dielectric Response law. Accordingly, the alternating current (AC) conductivity is a fractional power of frequency. The analysis of AC conductivity spectra recorded evidence for a proportionality between the logarithm of the pre-exponential factor to the fractional exponent in many cases. In the present work, the observed compensation is interpreted based on the two fundamental characteristics of disordered matter: universality and scaling. For this Ghosh and Pan Scaling Rule which links the direct current (DC) conductivity to the AC one, and the Jonscher Universal Dielectric Response merge to a single function of three quantities: DC conductivity, pre-factor and power exponent. Applying the partial differentiation chain theorem and taking the temperature dependencies of these quantities for Arrhenius systems, the compensation rule is derived.

Corresponding author: Anthony N. Papathanassiou (Email Address: antpapa@phys.uoa.gr)

## 1. Introduction

Disorder underlies scaling and universality of the alternating current (AC) conductivity in different classes of miscellaneous materials, such as amorphous semiconductors, ion or electron conducting polymers, granular metals, ion conducting glasses, composites, metal cluster compounds, metal oxides etc. [1-7]. Experimental results on a large number of materials belonging to different types of disordered media, different types of electric charge carriers (such as ions or electrons (polarons)) and different microscopic mechanisms (such as migration surpassing or phonon assisted tunneling of the potential barrier separating adjacent equilibrium sites) evidence for a common frequency dependent conductivity and common temperature dependency [1,2]. It is generally accepted that percolation causes and regulates the occurrence of AC universality and hence scaling [1,3]. The percolation network accessible to electric charge carriers emerge universality in direct current (DC) and AC conductivity in different types of materials. The electrical conductivity $\sigma_m(f)$ measured when applying an external harmonic electric field of frequency *f*, consists of two components: the macroscopic (DC) conductivity $\sigma_{dc}$ and the frequency-dependent AC one: $\sigma_{ac}(f)$. Over the past fifty years, numerous experimental results on a broad variety of disordered materials confirmed the Jonscher's Universal Dielectric Response (UDR) law, which refers to the universal shape of $\sigma_{ac}(f)$ isotherms [8, 9]. On the other hand, a Scaling Rule (SR) suggested by Ghosh and Pan [10], which employed the Almond and West function [11, 12], predicts quite successfully that different $\sigma_m(f)$ isotherms (of the same material) collapse to a single master curve, when frequency and conductivity axes are properly scaled. According to the Jonscher's Universal Dielectric Response (UDR), the AC conductivity as a function of frequency, at constant temperature *T*, obeys a fractional power relation [8,9]:

$$\sigma_{ac}(f,T) = A(T) \cdot f^{n(T)} \tag{1}$$

where $A(T)$ and $n(T)$ are a pre-exponential factor and a power exponent, respectively, which are determined by fitting eq. (1) to the $\sigma_{ac}(f,T)$ isotherms. The majority of experimental results of diverse materials indicate that $0 < n \leq 1$. However, supra-linear behavior ($n > 1$) has been observed in some cases [13,14].

Almond and West, trying to correlate DC and AC components in a defect mediated material with an activated number of charge carriers, proposed an analytic function for $\sigma_m(f,T)$ by defining $f_o$ the hopping rate for which the same number of hopping carriers contribute to the AC and DC conductivities, i.e., $\sigma_m(f_o,T) \equiv 2\sigma_{dc}$ [11]. Accordingly, an equation linking diffusive DC transport to sub-diffusive dispersive conductivity $\sigma_m(f,T)$ was employed to fit isothermal $\sigma_m(f,T)$ spectra [11,12,15]:

$$\sigma_m(f,T) = \sigma_{dc}(T) \cdot \left[1 + \left(\frac{f}{f_o(T)}\right)^N\right] \qquad (2)$$

The cross-over frequency $f_o$ signs the diffusive to sub-diffusive transition, while $N$ is a measure of ion motion correlations). In 2000, Ghosh and Pan [10] observed that hopping frequency can be used as the scaling frequency in the absence of well-defined dielectric loss peaks to scale conductivity spectra in ionic glasses for compositions of the same structure. This is justified adequately, since the change in hopping length with composition is manifested by the change in the hopping frequency resulting from correlation effects between successive hops. Frequency dependent conductivity data collected at different temperatures (for the same material) collapse on a master curve: $\frac{\sigma_m(f,T)}{\sigma_{dc}(T)}$ $vs$ $\frac{f}{f_o(T)}$ , when scaled quantities are plotted [1]. Ghosh and Pan [10], after physical reasoning, suggested a Scaling Rule (SR),by using a scaling function ***resembling*** that of Almond – West (Eq. (2)) to fit scaled $\frac{\sigma_m(f,T)}{\sigma_{dc}(T)}$ $vs$ $\frac{f}{f_o(T)}$ spectra of diverse materials quite successfully. Hereafter, according to the Ghosh and Pan Scaling Rule (SR), the exponent $N$ appearing in Eq. (2) is a single valued parameter characterizing the entire set of scaled isotherms [16-19]. N describes a set of AC conductivity isotherms, whereas n(T) describes individual isotherms. More precisely, *N* defines the overarching set of AC conductivity isotherms, whereas *n(T)* denotes the characteristics of individual isotherms.

Recasting Eq. (2) as: $\sigma_m(f,T) = \sigma_{dc}(T) + \sigma_{ac}(f,T)$ , $\sigma_{ac}$ $is$:

$$\sigma_{ac}(f,T) = \sigma_{dc}(T) \cdot \left(\frac{f}{f_o}\right)^N \qquad (3)$$

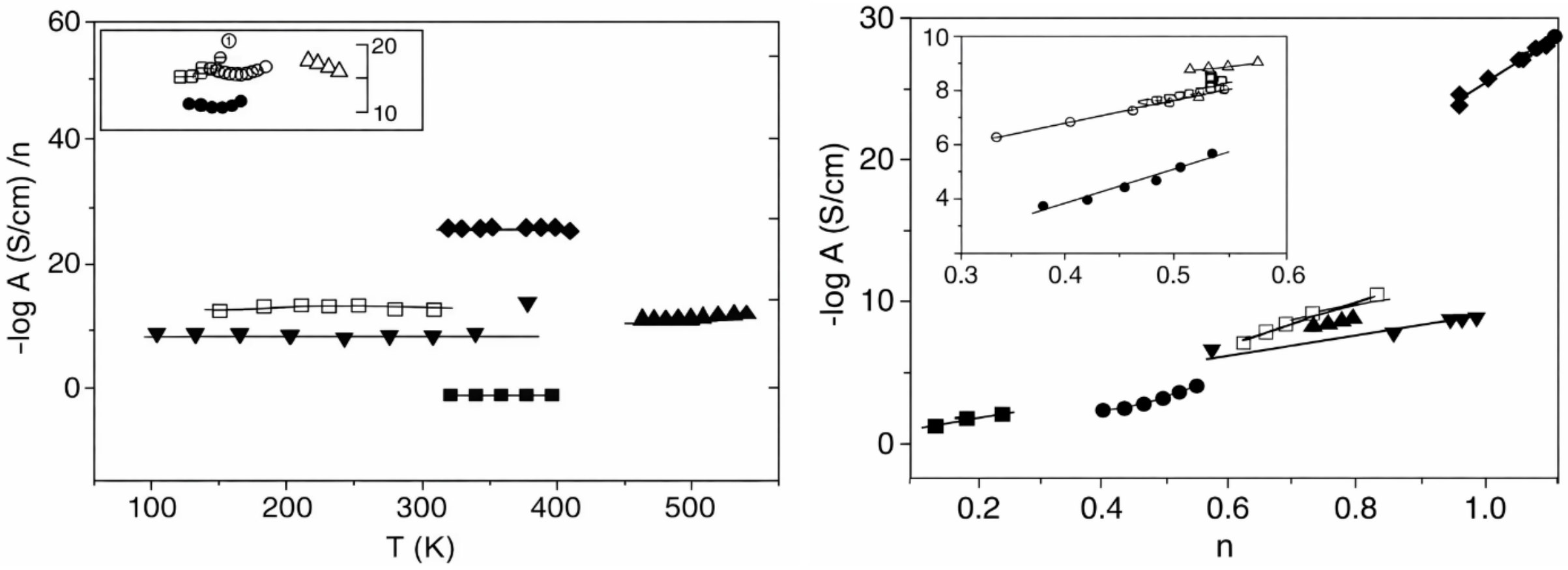

**Fig. 1.** Left plot: The ratio $-logA/n$ versus temperature for different types of materials: $(NH_4)_3H(SO_4)_{1.42}(SeO_4)_{0.58}$ (♦) Ref. [14]; $(NaNO_3–Al_2)_3$ (▲) Ref. [22]; $10CuI–60AgI–30V_2O_5$ (△) Ref. [23]; $xAg_2S–Sb_2S_3$; x = 70, 80 and 85, respectively (○, • and ⊕) Ref. [24]; $(CuI)_{0.55}–(AgSO_4)_{0.45}$ (■) Ref. [25]; $(Ag_2S)_{0.3}(AgPO_3)_{0.7}$ (□) Ref. [26]; $SeSm_{0.005}$ (▼) [27]. All data points share the same horizontal axis. Labels of the inset diagram are the same as those of the master one. Right plot: log A as a function of the exponent n of Eq. (1). Some data point sets are plotted in the inset diagram, in order to minimize overlap. Reproduced from [21] with permission from Elsevier.

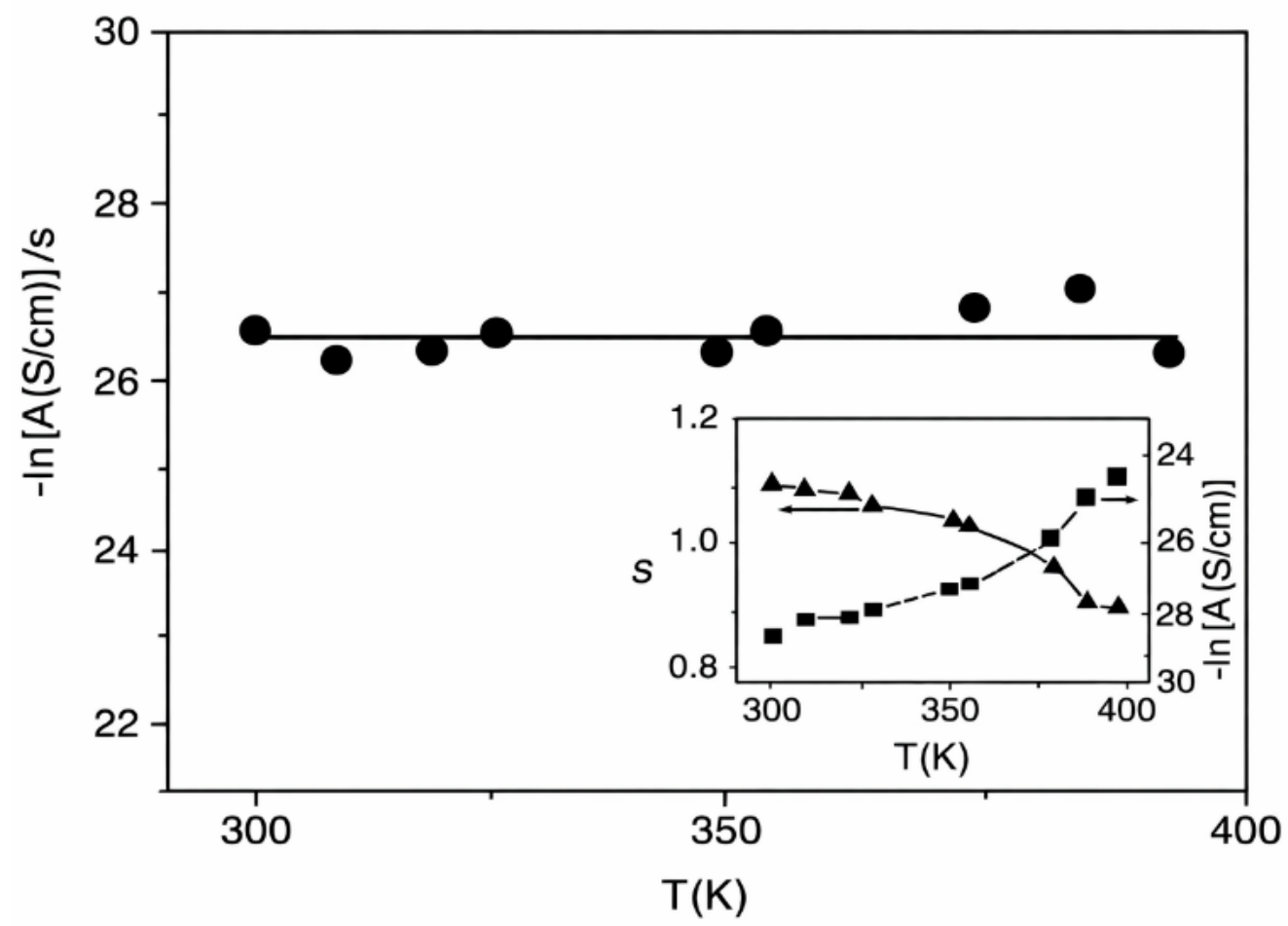

**Fig. 2.** The ratio $-lnA/s$ ($s$ denotes the slope $s \equiv dln\sigma_{ac}/dlnf$ and, hence, $s$=$n$) vs temperature of the mixed crystal $(NH_4)3H(SO_4)_{1.42}(SeO_4)_{0.58}$. In the inset, −lnA and s are plotted against temperature according to the results reported in [28]. Reproduced from [20] with permission from Elsevier.

## 2. Experimental evidence for a compensation rule

Except for the universality and scaling of electrical conductivity, a *compensation* between the parameters of the UDR law is observed [20, 21]: the values of $lnA$ and $n$ obtained by fitting Eq.(1) to AC conductivity spectra recorded at different temperatures, for the same solid, validate a proportionality relation: $lnA \propto n$, which implies that, in a $lnA$ vs $n$ diagram, data points lie on a single straight line, or the ratio $lnA/n$ is constant (independent upon temperature). In Fig 1, we can see that the ratios $logA/n$ vs temperature for $(NH_4)_3H(SO_4)_{1.42}(SeO_4)_{0.58}$ [14], $(NaNO_3–Al_2)_3$ [22], $10CuI–60AgI–30V_2O_5$ [23], $xAg_2S–Sb_2S_3$; x = 70, 80 and 85 [24], $(CuI)_{0.55}–(AgSO_4)_{0.45}$ [25], $(Ag_2S)_{0.3}(AgPO_3)_{0.7}$ [26] and $SeSm_{0.005}$ [27], are practically constant. When the same data points are plotted in a $logA$ vs $n$ plot (Fig. 1), straight lines can be fitted, confirming a compensation of $lnA$ to $n$. In Fig. 2, the values of the intercept and the exponent of the UDR (Eq. (1)) evidence for a compensation rule for the mixed crystal $(NH_4)3H(SO_4)_{1.42}(SeO_4)_{0.58}$ [28]. These early observations of a compensation law (lacking a theoretical interpretation) were strengthened from later and up to date literature: In Fig. 3, a compensation of $lnA$ to $n$ is observed for fused silica, whereas, a supra-linear UDR is observed [14] (note that our theoretical approach does not necessarily require the power exponent $n$ to be fractional. A compensation of $lnA$ to $n$ is detected for $0.8BiGd_{0.2}Fe_{0.8}O_3$-$0.2PbTiO_{0.2}$ [29] and sodium tantalate ($NaTaO_3$), doped with metal ions Cu (circles) and Al (squares) [30], respectively (Fig. 4), as well as, for pure cellulose and 4 wt % carbon nano-filler cellulose composite recorded at different temperatures, ranging from 20ºC to 100ºC [31] (Fig. 5). Recently, the transition from overlapping large polaron to small polaron tunneling induced on increasing temperature from 210 K to 417K was visualized through the temperature dependence of the ratio *lnA/n* for conduction mechanisms and phase transitions in long chain organic inorganic hybrids [32]. $lnA$ compensates to $n$ for $(Na_{0.33}Bi_{0.33}Sr_{0.33})(Ti_{1-x}W_{0.666x})O_3$ perovskite where $0 \leq x \leq 7$, for temperature ranging from 350 K to 450K [33] (Fig. 6). The paradigms mentioned above indicate that, for many different disordered materials (at least, for the systems reported above), compensation between the parameters *lnA* and *n* of the Jonscher's UDR occurs for data collected at different temperatures.

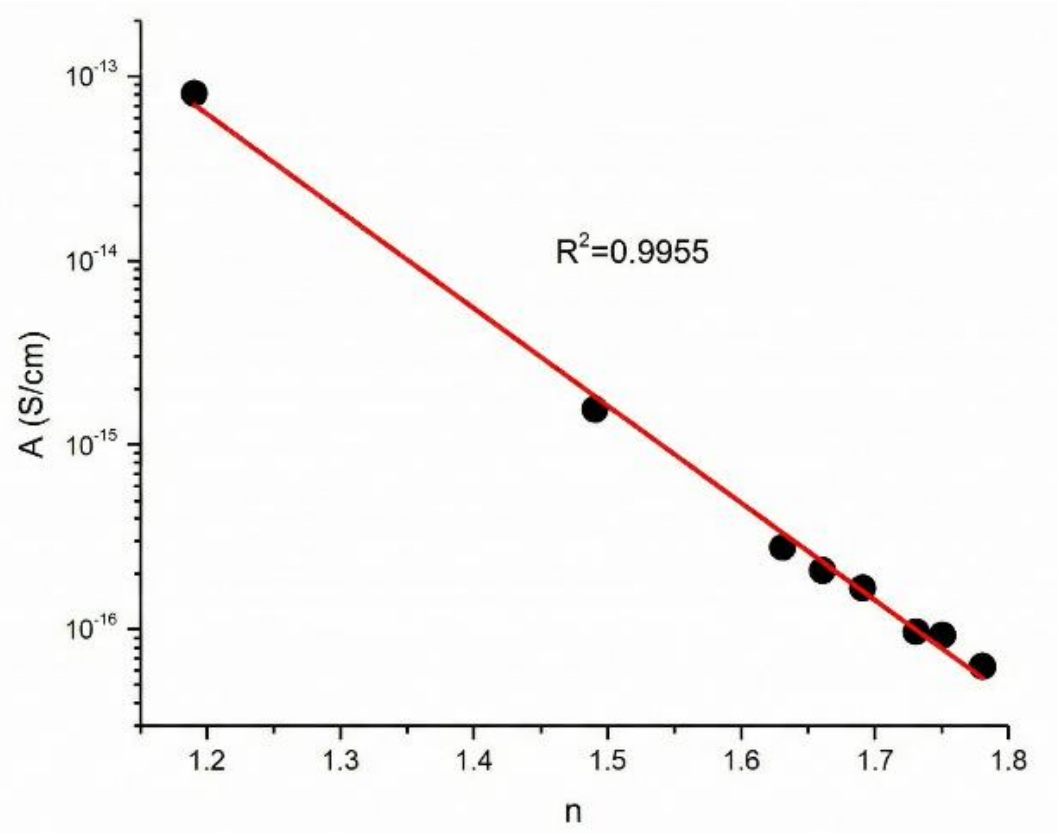

**Fig. 3.** Semi-logarithmic plot of A vs n from 250 to 700 ºC for fused silica [14]. Note that, even for the case of supra-linear UDR behavior, *lnA* compensates to *n*. $R^2$ denotes the coefficient of determination.

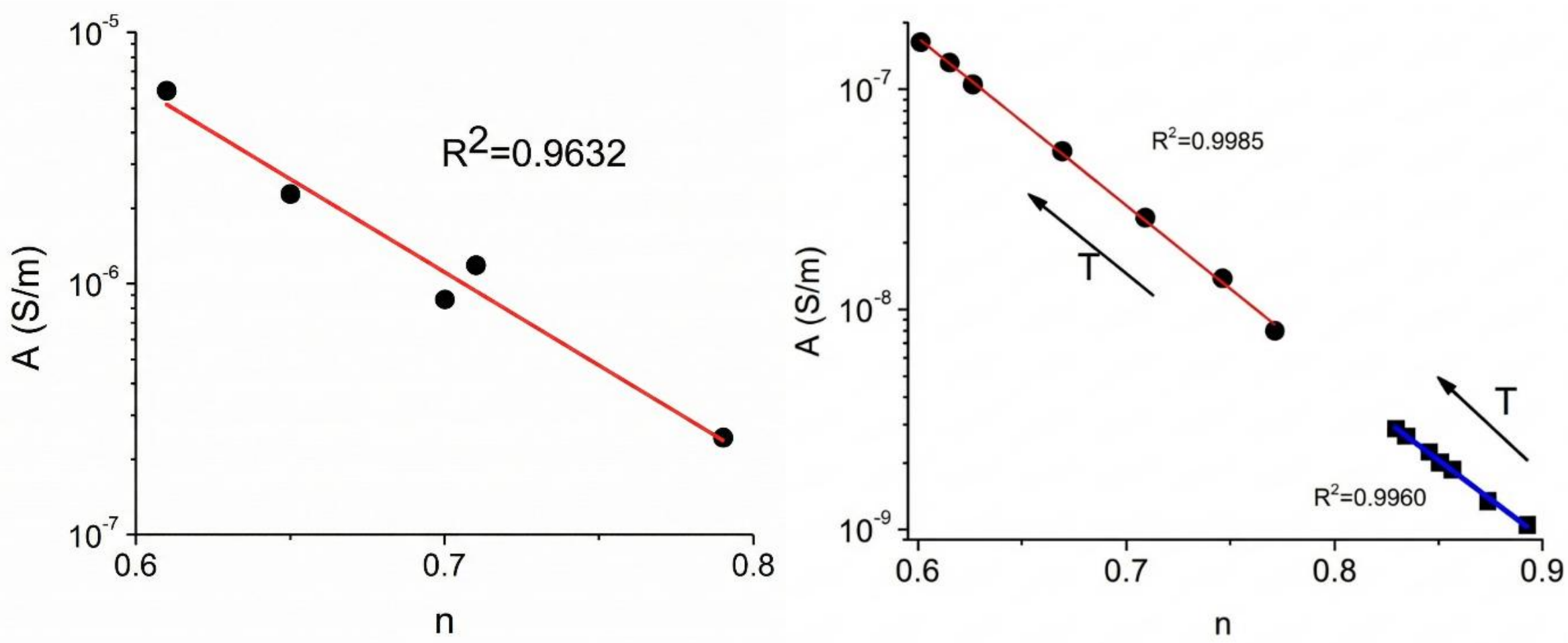

**Fig. 4.** Left diagram: The parameter *A* vs *n* of the UDR obtained from 250ºC to 350ºC for $0.8BiGd_{0.2}Fe_{0.8}O_3$-$0.2PbTiO_{0.2}$ [29]. Right plot: Logarithmic plot of the parameter *A* as a function of the fractional exponent *n* of the UDR (Eq. (1)) obtained from the analysis of the frequency - conductivity spectrum of sodium tantalate ($NaTaO_3$), doped with metal ions Cu (circles) and Al (squares), recorded at various temperatures ranging from 30ºC to 90ºC [30]. Straight lines fit best the data points, confirming a compensation of *lnA* to *n*. $R^2$ denotes the coefficient of determination.

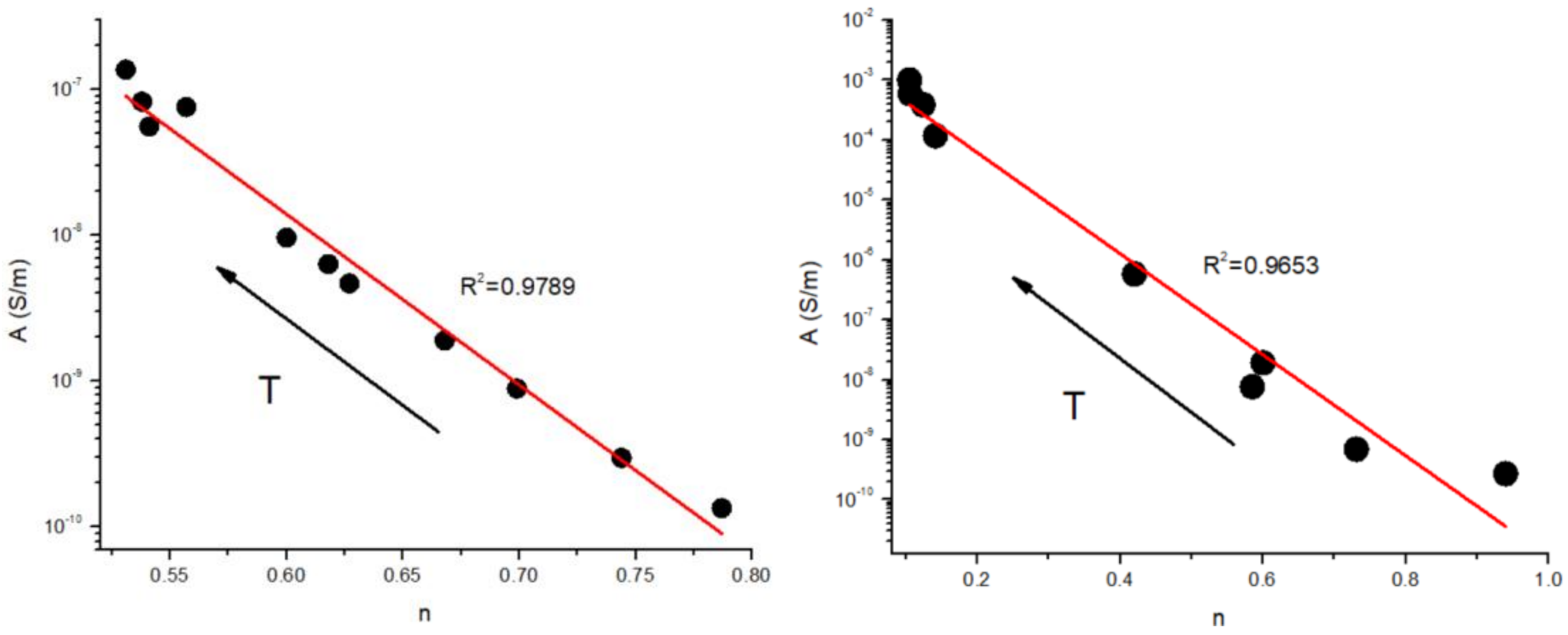

Fig. 5. Logarithmic plot of the parameter *A* as a function of the fractional exponent *n* of the UDR (Eq. (1)) obtained from the analysis of the frequency - conductivity spectrum of pure cellulose (left diagram) and 4 wt % carbon nano-filler cellulose composite (right plot), recorded at different temperatures ranging from 20ºC to 100ºC [31]. Straight lines best fit the data points, confirming a compensation of lnA to *n*. $R^2$ denotes the coefficient of determination.

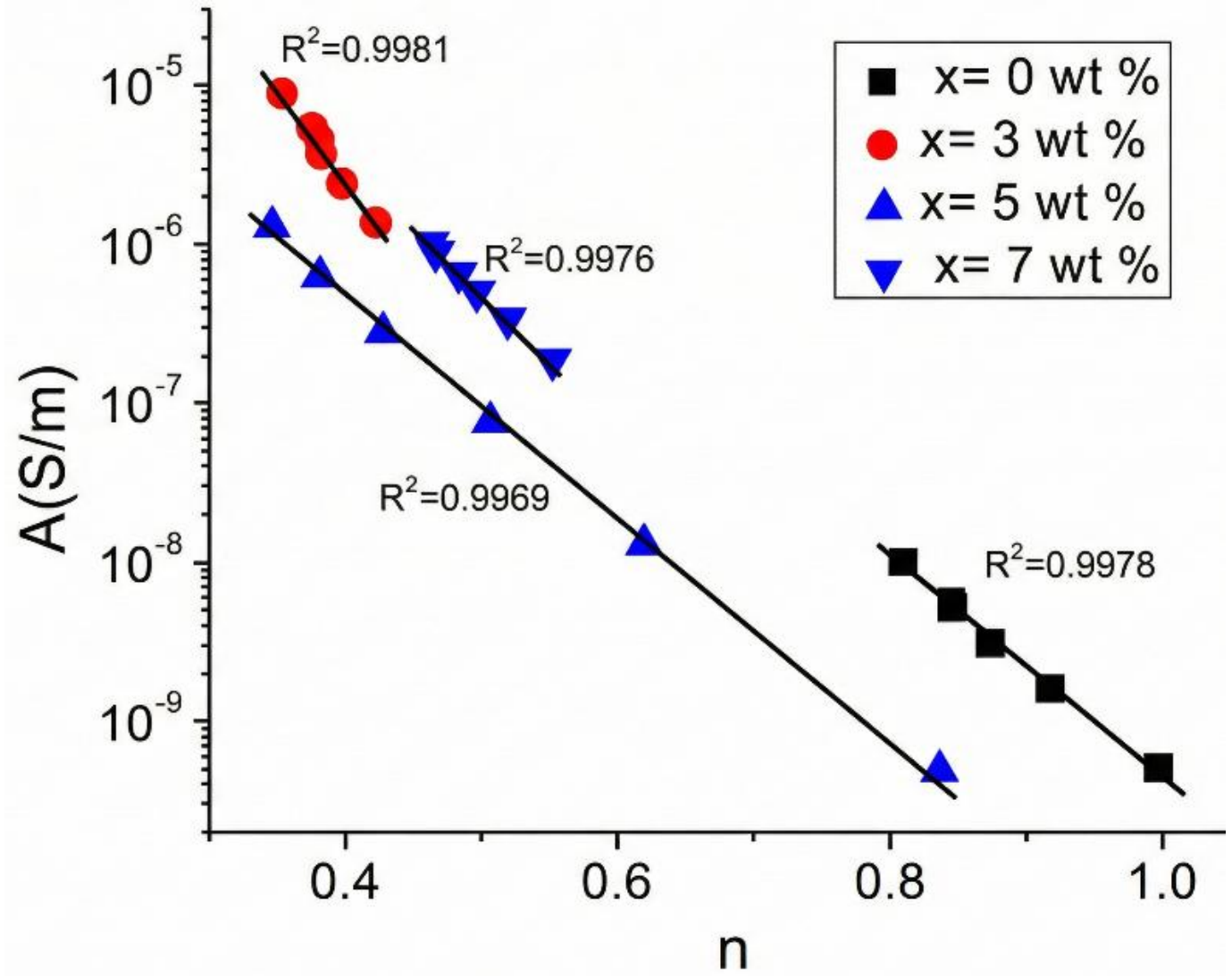

Fig. 6. Logarithmic plot of the parameter *A* as a function of the fractional exponent *n* of the UDR (Eq. (1)) obtained from the analysis of the frequency - conductivity isotherms ranging from 350 to 450 K for $(Na_{0.33}Bi_{0.33}Sr_{0.33})(Ti_{1-x}W_{0.666x})O_3$ perovskite, where $0 \le x \le 7$ [33]. Straight lines best fit the data points, confirming a compensation of *lnA* to *n*. $R^2$ denotes the coefficient of determination.

## 3. Theoretical interpretation

The origin of the observed *lnA* to *n* compensation will be addressed in this section, based on the two fundamental characteristics of disordered matter: UDR and SR, respectively. Electric charge transport and dielectric relaxation in disordered materials are characterized by a set of three quantities: $\sigma_{dc}$, $A$ and $n$, respectively. We recall that, in thermodynamic systems, the state parameters relate each other through an equation of state. Here, UDR (Eq. (1)) and SR (Eq. (3)) merge into an equation (analogous to the thermodynamic equation of state), which is a function of three quantities ($\sigma_{dc}$, $A$ and $n$):

$$\frac{\sigma_{dc}}{f_o^N} \cdot f^N = A \cdot f^n \tag{4}$$

The importance of Eq. (4) lies in capturing universality and scaling of the electrical and dielectric properties of disordered matter in a single function. A non-trivial solution of Eq. (4) is: $f = \left(\frac{\sigma_{dc}}{A f_o^N}\right)^{1/(n-N)}$. Eq. (4) is a composite function of $lnA(T)$, $n(T)$ and $ln\sigma_{dc}(T)$. Hence, the chain-rule theorem for three variables applies:

$$\left(\frac{\partial lnA}{\partial n}\right)_{ln\sigma_{dc}} \cdot \left(\frac{\partial n}{\partial ln\sigma_{dc}}\right)_{lnA} \cdot \left(\frac{\partial ln\sigma_{dc}}{\partial lnA}\right)_{n} = -1 \tag{5}$$

The physical meaning of the partial derivatives can be understood by examining, for example, one of them: $\left(\frac{\partial lnA}{\partial n}\right)_{ln\sigma_{dc}}$ indicates the explicit change $\Delta lnA$ upon a change $\Delta n$ induced by a temperature change $\Delta T$, regardless of the temperature change of $ln\sigma_{dc}$. The derivative $\left(\frac{\partial lnA}{\partial n}\right)_{ln\sigma_{dc}}$ can be written in a simple form: $\frac{dlnA}{dn} = \frac{dlnA/d(\frac{1}{kT})}{dn/d(\frac{1}{kT})}$, where $k$ denotes the Boltzmann constant. In turn, $\frac{dlnA}{dn}$ can be determined from the temperature dependencies of *lnA* and *n*. Eq. (5) gives:

$$\left(\frac{\partial lnA}{\partial n}\right)_{ln\sigma_{dc}} = -\left(\frac{\partial ln\sigma_{dc}}{\partial n}\right)_{lnA} \cdot \left(\frac{\partial lnA}{\partial ln\sigma_{dc}}\right)_{n} \tag{6}$$

We are going to examine whether the derivatives on the right-hand side of Eq. (6) are nearly constant.

Estimation of the term $\boldsymbol{\frac{dln\sigma_{dc}}{dn}}$***:*** The numerator of the derivative $\frac{dln\sigma_{dc}}{dn}$ describes the DC conductivity, while the denominator describes the AC conductivity. The DC conductivity corresponds to the diffusive dynamics, while the AC conductivity corresponds to the sub-diffusive dynamics. Nevertheless, the derivative $\frac{dln\sigma_{dc}}{dn}$ is physically meaningful as DC and AC dynamics are correlated with each other through: (i) scaling functions (such as Eq. (2)), or (ii) the Barton–Nakajima–Namikawa (BNN) relation: $\sigma_{dc} = p\varepsilon_0 \Delta\varepsilon 2\pi f_m$, where $\varepsilon_0$ denotes the permittivity of vacuum, $\Delta\varepsilon$ is the intensity of a dielectric loss peak maximizing at frequency $f_m$, associated with the transition from the DC regime to the dispersive AC conductivity (recall that the dielectric loss is expressed through the imaginary part of the complex (relative) permittivity: $Im(\varepsilon^*) = \frac{\sigma_m(f)-\sigma_{dc}}{2\pi\varepsilon_0 f}$) and $p$ is a numerical constant of the order of unity) [35 - 37] The dielectric loss is $Im(\varepsilon^*) = \frac{\sigma_m(f)-\sigma_{dc}}{2\pi\varepsilon_0 f}$, where $Im(\varepsilon^*)$ is the imaginary part of the complex (relative) permittivity, which maximizes at a frequency $f_m \approx f_0$. Let us now estimate the term $\frac{dln\sigma_{dc}}{dn}$ by considering how $ln\sigma_{dc}$ and $n$ evolve with temperature:

$$\frac{dln\sigma_{dc}}{dn} = \frac{dln\sigma_{dc}/d(\frac{1}{kT})}{dn/d(\frac{1}{kT})} \tag{7}$$

where $k$ is the Boltzmann's constant and $T$ is the absolute temperature. If the DC conductivity obeys an Arrhenius relation (actually, an approximate Arrhenius law can be used to describe $\sigma_{dc}(T)$ in non-Arrhenius systems, for a narrow temperature range), and activation energy is defined as:

$$E_{dc} \equiv -dln\sigma_{dc}/d(\frac{1}{kT}) \tag{8}$$

The temperature dependence of the power exponent $n(T)$ depends on the material and the microscopic AC conductivity process. Most of experimental data conclude that $n(T)$ is an decreasing function: from a fractional value at room temperature to unity when approaching low temperatures (e.g., $n(T = 77\ K) \rightarrow 1)$) [38]. Exceptions for different $n(T)$ evolutions have been observed, as a result of material's properties and/or the experimental inaccuracy in determining $n(T)$ due to limited temperature range, sparce temperature sampling and limited number of $n(T)$ data points on recording AC conductivity isotherms. For carbon nanotubes-polyepoxy composites [42], bulk furfurylidenemalononitrile [40] or chalcogenide glassy system, xAg2S-(1−x) (0.5Se–0.5Te) with x = 0.05, 0.1 and 0.2 [40] measured from room temperature to 423 K, 333 K and 383 K, respectively, a constant negative value of $\frac{dn}{d(\frac{1}{kT})}$ was reported, in agreement with the predictions of the Quantum Mechanical Tunneling model [38]. Correlated Barrier Hopping has interpreted numerous experimental findings reporting a decreasing function of $n(T)$ [39, 41]. For large binding energy $w$ of the electric charge carrier to its localized state (i.e., $w \gg kT$) which is related to the maximum barrier height at infinite inter-site separation, $n(T)$ decreases linearly upon $T$: $n(T) = 1 - cT$, where $c \equiv 6k/w$ is a constant. Given that $0 < cT < 1$, equation $n(T) = 1 - cT$ reads: $cT = 1 - n$ or $\frac{1}{cT} = \frac{1}{1-n}$ . To a first approximation, by keeping first order terms of the geometric expansion of the term $\frac{1}{1-n}$, we get: $\frac{1}{1-n} \approx 1 + n$. Subsequently, $\frac{1}{cT} \approx 1 + n$, which gives: $\frac{dn}{d(\frac{1}{kT})} \approx \frac{k}{c}$.

While our mathematical treatment appears to be well suited the Quantum Mechanical Tunneling (QMT)

and Correlated Barrier Hopping (CBH) models (whereas n(T) is a decreasing near-Arrhenius function), its relevance to the Overlapping Large Polaron Tunneling (OLPT) model and the Non Overlapping Small Polaron Tunneling (NSPT) model remain uncertain. A source of such limitation is likely to occur from the specific temperature dependence of the fractional exponent: n increases with temperature I\in NSPT model, while n(T) changes its monotony in OLPT. We note that the temperature evolution of $n(T)$ regulates the accuracy of the predictions of our theoretical approach and may possibly account for cases where compensation is hard to observe. Discrepancies may result either from the non-Arrhenius dependence of $n(T)$ or from uncertainties arising when few experimental points are available. For Arrhenius type materials obeying the approximate condition $\frac{dn}{d(\frac{1}{kT})} \cong constant$, Eq. (7) writes:

$$\frac{dln\sigma_{dc}}{dn} \cong constant \tag{9}$$

Estimation of the term $\frac{dlnA}{\partial dln\sigma_{dc}}$ *:* The pre-exponential factor *A(T)* of the UDR (Eq. (1)) coincides with $\sigma_{ac}(f \to 0, T)$ in the limit of zero frequency: $lnA(T) = ln\sigma_{ac}(f \to 0, T)$. Assuming an Arrhenius temperature dependence of $\sigma_{ac}(f \to 0, T)$, an activation energy for AC conductivity is defined as: $E_{ac}(f \to 0) \equiv -\frac{dln\sigma_{ac}(f \to 0, T)}{d(\frac{1}{kT})}$. Subsequently:

$$-\frac{dlnA(T)}{d(\frac{1}{kT})} = E_{ac}(f \to 0) \tag{10}$$

Due to Eqs. (8) and (10), the first term of the product of the right-hand side of Eq. (6) reads:

$$\left(\frac{\partial lnA}{\partial ln\sigma_{dc}}\right)_n = \left(\frac{\partial lnA}{\partial(\frac{1}{kT})}\right)_n \cdot \left(\frac{\partial(\frac{1}{kT})}{\partial ln\sigma_{dc}}\right)_n = E_{ac}(f \to 0)/E_{dc} \tag{11}$$

The fraction $E_{ac}(f \to 0)/E_{dc}$ is constant, as a ratio of temperature independent quantities. To a first approximation, $E_{ac}(f \to 0)/E_{dc}$ is the ratio of the lowest to the highest effective energy barriers of the energy landscape within a disordered material. Thus, $\left(\frac{\partial lnA}{\partial ln\sigma_{dc}}\right)_n$ is a measure of the width of the distribution of potential energy barriers and characterizes the degree of disorder of a material. Eq. (11) writes:

$$\left(\frac{\partial lnA}{\partial ln\sigma_{dc}}\right)_n \cong constant \tag{12}$$

Within certain assumptions and approximations mentioned above, we found that the right-hand side of Eq. (6) is nearly constant, as a product of two nearly constant terms (Eqs. (9) and (12): $\frac{dlnA}{dn} \cong$ $constant$. Alternatively, $lnA$ is proportional to $n$ (i.e., $lnA$ compensates with $n$). Such compensation rule applies to systems characterized by an Arrhenius behavior in the DC and AC conductivities (Eqs. (8) and (10), respectively). To understand possibly why the present theory applies to Arrhenius systems, we have to recall that physical meaning of the fraction $E_{ac}(f \to 0)/E_{dc}$ lies in the fact that it is a measure of the width of the distribution of the potential barrier heights. A constant value of $E_{ac}(f \to 0)/E_{dc}$, implies that the potential height distribution does not change with temperature emerging scaling, universality and compensation of AC conductivity measured at different temperatures (for the same material). Nevertheless, even for non-Arrhenius systems, if AC conductivity spectra are collected in a narrow temperature range, a quasi-Arrhenius behavior can be assumed, and compensation rule may also apply. Moreover, compensation requires that $dn/d(\frac{1}{kT})$ retains a non-zero constant value, which is approximately valid for linear dependencies of $n(T)$. Our work complies with a model $n(T) = 1 - (T/T_0)$ ($T_0$ denotes a constant) an approximately linear dependence of$lnA$ on temperature (or on $n$) is obtained [43].

A final, interesting question that remains to be addressed is whether the product $\left(\frac{\partial lnA}{\partial n}\right)_{ln\sigma_{dc}} \cdot \left(\frac{\partial n}{\partial ln\sigma_{dc}}\right)_{lnA} \cdot \left(\frac{\partial ln\sigma_{dc}}{\partial lnA}\right)_{n}$ results in a negative as predicted by Euler's chain rule. According to Eq. (7): $\frac{dln\sigma_{dc}}{dn} = \frac{dln\sigma_{dc}/d(\frac{1}{kT})}{dn/d(\frac{1}{kT})}$. Per Eq. (8), the nominator equals $-E_{dc}$, which is inherently negative. Meanwhile, the literature demonstrates that *n(T)* is a decreasing function of temperature for a wide range of materials, meaning that *n* increases as $\frac{1}{kT}$ increases. Therefore, the denominator $dn/d(\frac{1}{kT})$ is positive. A negative numerator divided by a positive denominator mathematically ensures that $\frac{dln\sigma_{dc}}{dn} < 0$ , or: $\left(\frac{\partial n}{\partial ln\sigma_{dc}}\right)_{lnA} < 0$. Besides, according to Eq. (11), $\left(\frac{\partial lnA}{\partial ln\sigma_{dc}}\right)_{n} = \left(\frac{\partial lnA}{\partial(\frac{1}{kT})}\right)_{n} \cdot \left(\frac{\partial(\frac{1}{kT})}{\partial ln\sigma_{dc}}\right)_{n} = E_{ac}(f \to 0)/E_{dc}$ >0. Finally, the experimental results presented in the present work indicate that the term $\left(\frac{\partial lnA}{\partial n}\right)_{ln\sigma_{dc}}$ is positive. We conclude that the product $\left(\frac{\partial lnA}{\partial n}\right)_{ln\sigma_{dc}} \cdot \left(\frac{\partial n}{\partial ln\sigma_{dc}}\right)_{lnA} \cdot \left(\frac{\partial ln\sigma_{dc}}{\partial lnA}\right)_{n}$ appearing in Eq. (5) is negative, indeed, in compliance with chain rule.

## 4. Conclusion

Experimental results on different classes of disordered solids evidence for a compensation of the logarithm of the pre-exponential factor *lnA* to the power exponent *n* of the Universal Dielectric Response (UDR). A theoretical framework for a compensation rule between parameters of Jonscher's UDR law in disordered materials, within certain assumptions and approximations, is developed in the present work.

DC conductivity, pre-exponential factor and fractional exponent of AC conductivity are three fundamental quantities, which regulate the electrical and dielectric properties of a material. These constitute a unique equation, which is derived by merging the UDR law and the Ghosh – Pan Scaling Rule (SR). A partial differentiation chain theorem is applied. Accordingly, partial derivatives are estimated from the temperature dependencies of the DC conductivity, pre-exponential factor and fractional exponent of the AC response, yielding a compensation rule between the parameters of the UDR law. The compensation rule applies to systems characterized by an Arrhenius behavior of the DC and AC conductivities (see Eqs. (8) and (10), respectively). However, even for non-Arrhenius systems, if AC conductivity spectra are collected within a narrow temperature range, a quasi-Arrhenius behavior can be assumed, and compensation rule may also apply. Moreover, compensation requires that $dn/d(\frac{1}{kT})$ retains a non-zero constant value, which is approximately valid if $n(T)$ is linear. Interestingly, a compensation rule between the parameters of the UDR was reported by varying the composition of mixed alkali halide crystals at constant temperature, $lnA$ compensates to $n$, as well [34] and poses raises questions for further investigation: The numerators and denominators of the derivatives appearing in the chain rule (Eq. (4)) could be expressed as a function of concentration or pressure (instead of temperature) to derive a compensation rule for concentration or pressure AC conductivity experiments. In this sense, we believe that the theoretical frame of the present work holds potential for unifying AC conductivity descriptions in disordered systems.

### Author contributions (CRediT)

A.N.P. contributed conceptualization, data curation, formal analysis, investigation, methodology, validation, visualization and writing. E.S. read the manuscript critically.

### Data availability statement

Data is available on request from the authors.

### Funding

No funding.

### Ethics

Ethics, Consent to Participate, and Consent to Publish declarations: not applicable.

### Competing interest declaration

The authors have no relevant financial or non-financial interests to disclose. The authors have no conflicts of interest to declare that are relevant to the content of this article. All authors certify that they have no affiliations with or involvement in any organization or entity with any financial interest or non-financial interest in the subject matter or materials discussed in this manuscript. The authors have no financial or proprietary interest in any material discussed in this article.